\documentclass[twocolumn,nofootinbib,superscriptaddress]{revtex4-1}
\usepackage{graphicx,epsfig}
\usepackage{amsmath}
\usepackage{amsfonts}
\usepackage{braket}
\usepackage{fancyhdr}
\usepackage{cleveref}
\usepackage[utf8]{inputenc}
\usepackage{color}
\newcommand{\be}{\begin{eqnarray}}
\newcommand{\ee}{\end{eqnarray}}

\newcommand{\com}{{\cal C}}
\newcommand{\cc}{\bar\com_{\rm c}}

\begin{document}





\title{A universal threshold for primordial black hole formation}
\author{Albert Escriv\`a}
\email{albert.escriva@fqa.ub.edu}
\affiliation{Institut de Ci\`encies del Cosmos, Universitat de Barcelona, Mart\'i i Franqu\`es 1, 08028 Barcelona, Spain}
\affiliation{Departament de F\'isica Qu\`antica i Astrof\'isica, Facultat de F\'isica, Universitat de Barcelona, Mart\'i i Franqu\`es 1, 08028 Barcelona, Spain}
\email{albert.escriva@fqa.ub.edu}
\author{Cristiano Germani}
\email{germani@icc.ub.edu}
\affiliation{Institut de Ci\`encies del Cosmos, Universitat de Barcelona, Mart\'i i Franqu\`es 1, 08028 Barcelona, Spain}
\author{Ravi K. Sheth}
\email{shethrk@upenn.edu}
\affiliation{Center for Particle Cosmology, University of Pennsylvania, Philadelphia, PA 19104, USA}

\begin{abstract}
	In this letter, we argue and show numerically that the threshold to form primordial black holes from an initial spherically symmetric perturbation is, to an excellent approximation, universal, whenever given in terms of the compaction function averaged over a sphere of radius $r_m$, where $r_m$ is the scale on which the compaction function is maximum. This can be understood as the requirement that, for a black hole to form, each shell of the averaged compaction function should have an amplitude exceeding the so-called Harada-Yoo-Kohri limit.  For a radiation dominated universe we argued, supported by the numerical simulations, that this limit is $\delta_c = 0.40$, which is slightly below the one quoted in the literature.  Additionally, we show that the profile dependence of the threshold for the compaction function is only sensitive to its curvature at the maximum. We use these results to provide an analytic formula for the threshold amplitude of the compaction function at its maximum in terms of the normalised compaction function curvature at $r_m$.
\end{abstract}

\maketitle

\section{Introduction}

In a Friedmann-Robertson-Walker (FRW) universe filled with a single fluid component having equation of state $p=\omega\rho$, a spherically symmetric local perturbation can be approximated, to leading order in gradient expansion (super-horizon scales), as\footnote{The metric is gauge dependent. Nevertheless, at leading order in gradient expansion, several gauges give the same result. To fix ideas we considered the Kodama-Sasaki gauge \cite{sasaki_gauge}.}
\be
ds^2\simeq-dt^2+a(t)^2\left[\frac{dr^2}{1-K(r)r^2}+r^2d\Omega^2\right]\ .
\ee
Here, the local ``gravitational potential'' ($K(r)r^2$) parameterizes the initial curvature perturbation. We assume in the rest of the paper that the relevant fluctuations are all at super-horizon scales.

In \cite{yokoyama}, numerical simulations were used to argue that the threshold for the amplitude of an over-density peak forming a spherically symmetric black hole in a FRW universe, only depends upon two master parameters: the integral of the initial $K(r)$ and the edge of the over-density distribution. 
Musco \cite{musco} recently refined the arguments of \cite{yokoyama} by showing that the threshold may be more conveniently given in terms of the amplitude of the gravitational potential at its maximum ($r=r_m$), as already noticed in \cite{shibata}, and that it mainly depends upon the functional form (shape) of the gravitational potential up to $r_m$. More precisely, in \cite{musco}, the threshold was given in terms of the ``compaction function'' \cite{shibata} at super-horizon scales $\com(r)$ (here and after we shall simply call it compaction function)\footnote{The definition in \cite{shibata} differs by a factor $2$ with respect to the one used here.}:  The compaction function, which closely resembles the Schwarzschild gravitational potential, is defined as twice the local excess-mass over the co-moving areal radius.  At super-horizon scales it is (in units $G_N=1$)
\be
 \com(r) = f(\omega)K(r)r^2\ ,
\ee 
where $f(\omega)\equiv 3\left(1+\omega\right)/\left(5+3\omega\right)$. From this, one finds $r_m$ as the first root of $\com'(r)=0$. For a radiation dominated universe, $\omega=1/3$ and so $f(\omega)\equiv 2/3$.

Regularity -- the gravitational potential within a vanishingly small volume must be zero -- ensures that $K(r)r^2\rightarrow 0$ for $r\rightarrow 0$. Thus, the behavior of $K(r)$ around the origin plays little role in black hole formation.
In addition, the threshold for primordial black hole formation should be quite insensitive to the behaviour beyond $r_m$ as already numerically noticed in \cite{musco}. The reason is simple: the threshold is the amplitude above which a ``virtual'' black hole of zero mass is formed. Therefore, all the over-density beyond $r_m$ will be diffused away while that just in the vicinity of $r_m$ will hinder collapse. Hence, we also expect the threshold to be very weakly dependent on the exact scale of $r_m$.  
However, above threshold, a larger part of the initial profile would be involved in the collapse. This is why the mass of the black hole with peak density above threshold also depends non-trivially on the scale $r_m$ on which the compaction function peaks. 

Summarizing, while we expect the threshold to depend on profile shape, this dependence should come mainly from $\com$ around $r_m$.  Since $r_m$ is the scale on which $\com'=0$, we expect the threshold to depend primarily on $\com''$ at $r_m$. We explore this further in the next section.

\section{Approximating the curvature}

We have checked numerically that the family of centrally peaked exponentials used in \cite{musco}, for the purpose of obtaining the threshold for black hole formation, is an efficient basis with which to approximate any compaction function around its maximum.

By defining the parameter
\be\label{q}
 q\equiv-\frac{\com''(r_m)r_m^2}{4\,\com(r_m)}
\ee
we consider the basis 
\be\label{basis}
 K_b(r)=\frac{\com(r_m)}{f(\omega)r_m^2}e^{\frac{1}{q}\left(1-\left[\frac{r}{r_m}\right]^{2q}\right)}\ .
\ee
Note that $K_b(r) \to\com(r_m)/f(\omega)r_m^2 \theta(r_m-r)$ as $q\to\infty$.  This `homogeneous sphere' limit will be useful below.  In contrast, $K_b\to \com(r_m)/f(\omega)r^2$ and so $C(r)\to C(r_m)$ as $q\to 0$. 
In addition, note that $r_m$ only defines the units of length for the scaling of $K_b$. Thus, as already argued before, although the mass depends on $r_m$, the threshold value, defined at super-horizon scales, does not and, it is constant in time\footnote{ Of course this is only strictly correct at leading order in gradient expansion. At the full non-linear level the threshold would depend upon time. At super-horizon scales, where we put the initial conditions for the gravitational collapse, however, this dependence is negligible.}.

We have tested our basis by considering a representative class of curvatures for the case of a radiation dominated universe ($\omega=1/3$). We have found that the threshold for black hole formation, obtained by the use of our basis, only differs by $(1\div 2)\%$ from the one obtained by the exact curvature profiles considered. 

In the next section we use the basis \eqref{basis} to provide an analytical formula for the thresholds. We demonstrate that our formula accurately reproduces the numerical results obtained from the publicly available code for black hole formation of \cite{albert}. In turn, this will also show numerically our claim that the basis \eqref{basis} well-approximates any realistic desired curvature for the calculation of the threshold. 

\section{Universal threshold}

As noticed by \cite{musco}, the threshold for $\com(r_m)$ is not universal: it depends upon the shape of the curvature profile. This implies that, if initial conditions for primordial black hole (PBH) formation are generated during inflation (see e.g. \cite{vicente,inf}), then the threshold for PBH formation strongly depends on the form of the inflationary power spectrum \cite{germani}. For example, for an almost gaussianly distributed over-density field, the mean profile around a rare peak is proportional to the two-point correlation function of the field and hence, the power spectrum \cite{bbks,germani}\footnote{
		In the estimates of PBH abundances by the use of peak theory in over-densities (e.g. \cite{germani}), one needs a threshold for the central amplitude of the un-smoothed over-density, rather than $\com(r_m)$ (see however \cite{sg} for a non-linear approach).  
		The relation between these two quantities has been worked out in \cite{vicente} and only depends on the power spectrum of curvature perturbations calculated with linear analysis.}. Similarly, the parameter $q$, which will give us information about the {\it non-linear} evolution of the perturbation, only depends upon the details of the inflationary power spectrum and thus is solely related to the {\it linear} analysis of inflation. 

What we show below is the remarkable fact that, nevertheless, the threshold for the average compaction function is, within $2\%$ with respect to the simulations, {\it universal}.  

Let us define
\be
 \bar\com_{\rm c}\equiv\frac{3}{r_m^3}\int_0^{r_m} \com_c(x) x^2 dx\ ,
\ee
where $\com_c(r)$ is the critical compaction function for generating a black hole with zero mass. By using the basis \eqref{basis}, we have
\be\label{cc}
 \cc=\frac{3}{2}e^{\frac{1}{q}}q^{-1+\frac{5}{2q}}\left[\Gamma\left(\frac{5}{2q}\right)-\Gamma\left(\frac{5}{2q},\frac{1}{q}\right)\right] \delta_c ,
\ee
where $\Gamma(x)$ is the gamma function, $\Gamma(x,y)$ the incomplete gamma function and we have defined\footnote{Note that in \cite{musco} $\delta_c$ is $\delta_m$.} $\delta_c\equiv\com_c(\rm r_m)$. Note that $\delta_c=3\frac{\delta\rho}{\rho}\frac{r_m^2}{a^2H^2}$ at super-horizon scales \cite{musco}, where $\frac{\delta\rho}{\rho}$ is the over-density and $H$ is the Hubble expansion.

Following \cite{musco}, if the initial perturbation is not already a black hole, the compaction function is bounded by $f(\omega)$. When radiation dominates then $\delta_c\to f(\omega)=2/3$ as $q\rightarrow\infty$, as shown numerically in \cite{musco}.  At large $q$ we have $\cc\sim (3/5)\,\delta_c$ so $\cc=2/5$. Our assumption, that we will prove both numerically and argue in the following, is that
\be\label{threshold}
 \cc = \frac{2}{5}
\ee
for {\it any} value of $q$.  (E.g., this implies that when $q\to 0$ then $\delta_c\to 2/5$.)  More generally, we find that the threshold for different curvature profiles $K(r)$ (when radiation dominates) is given by 
\be\label{deltac}
 \delta_c=\frac{4}{15} e^{-\frac{1}{q}}\frac{q^{1-\frac{5}{2q}}}{\Gamma\left(\frac{5}{2q}\right)-\Gamma\left(\frac{5}{2q},\frac{1}{q}\right)}\ ,
\ee
where $q$ is given by \eqref{q}.

\subsection{Numerical checks}

In this section, we use the publicly available code developed in \cite{albert}, to check the accuracy of \eqref{threshold} and \eqref{deltac}. 

\begin{figure}[b]
	\includegraphics[width=0.9\linewidth]{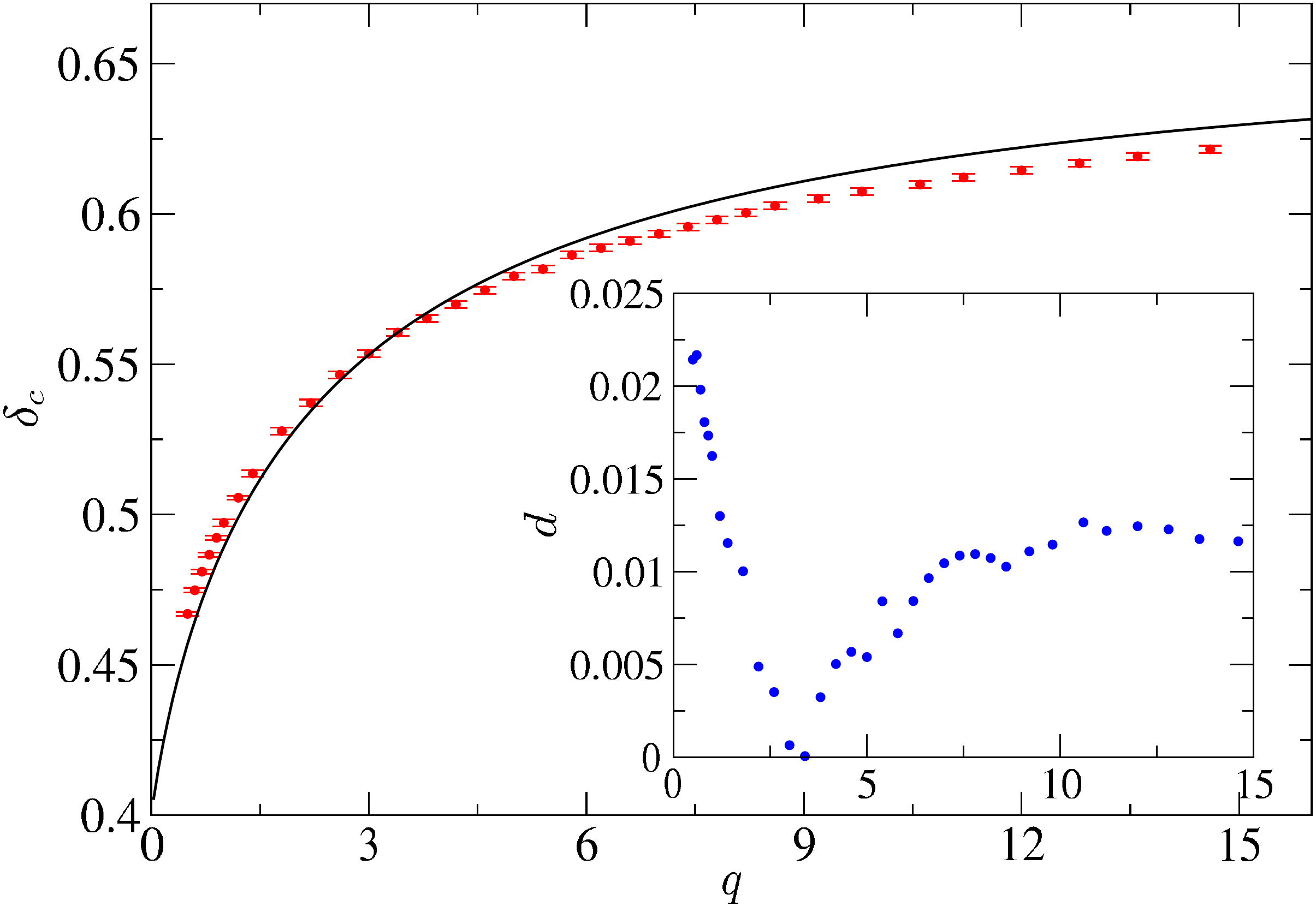} 
	\caption{Dependence of threshold on curvature $q$ (Eq.~\ref{q}) for the exponential basis profile $K_{b}$ (Eq.~\ref{basis}). Black curve shows $\delta_{c}^{A}$ (Eq.~\ref{deltac}); red points show $\delta_{c}^{N}$ obtained from simulations.}
	\label{fig:exponential}
\end{figure}

First, Fig.~\ref{fig:exponential} shows that \eqref{deltac} (solid curve) provides a good description of the threshold $\delta_{c}^{N}$ measured in the simulations (symbols) for profiles parametrized by \eqref{basis}.  The subpanel shows the relative errors $d\equiv|\delta_{c}^{N}-\delta_{c}^{A}|/\delta_{c}^{N}$: the agreement is better than $\sim 98\%$ for all $q$.  

To show that our results are more general than \eqref{basis}, we have also considered the following families for $K(r)$:
\begin{align}
 \label{eq:pol}
 K_{1} &=\frac{3}{2}\frac{\com(r_m)}{r_m^2}\frac{p/(p-2)}{1+\frac{2}{p-2}\left(\frac{r}{r_{m}}\right)^{p}} ; \\
 \label{eq:lamda}
 K_{2} &= \frac{3}{2}\frac{\com(r_m)}{r_m^2}\,
        \left(\frac{r}{r_{m}}\right)^{2\lambda}\,
        e^{\frac{1+\lambda}{\alpha}\left(1 - \left(\frac{r}{r_{m}}\right)^{2\alpha}\right)} ; \\
 \label{eq:spectrum}
 K_{3} &= \frac{3}{2}\frac{\com(r_m)}{r_m^2}\frac{r_m^3}{r^3}\,
         \frac{g(n,k_p,r)}{g(n,k_p,r_m)} ,
\end{align}
where
\begin{align}
 g(n,k_p,r) &= \left[k_{p}r \left\{ E_{3+n}(-ik_{p}r)+E_{3+n}(ik_{p}r)\right\} \right. \nonumber \\
&\ + i\left. \left\{-E_{4+n}(ik_{p}r)+E_{4+n}(-ik_{p}r) \right\} \right] ,
\end{align}
with $E_{n}(x) \equiv \int_{1}^{\infty}e^{-x t}\,dt/t^{n}$.

\begin{figure}[t]
	\includegraphics[width=0.8\linewidth]{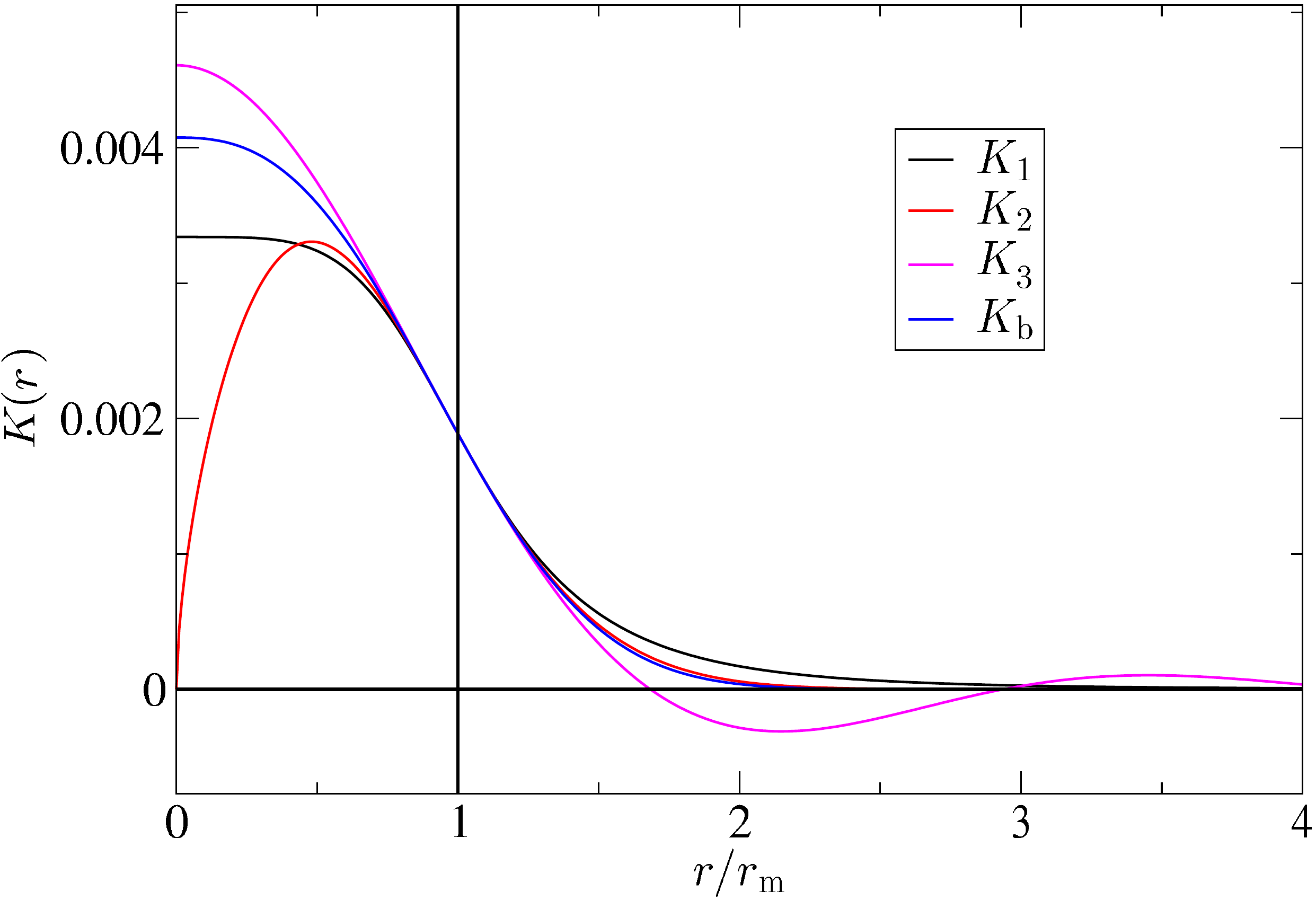} 
	\caption{Comparison between the profiles listed in eqs. \ref{eq:pol}-\ref{eq:spectrum} and $K_b$ in the case of the same threshold (proportional to the curvature value at $r=r_m$). For illustration, we have chosen the case $q=1.3$ leading to $\delta_{c} \approx 0.5035$. For $K_{1}$, this translates to $p=4.6$; for $K_{2}$ one has $\alpha=1$ and $\lambda=0.3$; for $K_{3}$, $n\approx 6.67$.}
	\label{fig:curvatures}
\end{figure}

Fig.~\ref{fig:curvatures} illustrates that $K_b$ is able to provide a good approximation to \eqref{eq:pol}, \eqref{eq:lamda} and \eqref{eq:spectrum} around $r_m$, for a few representative parameter choices.  The (oscillating) profile, $K_3$, is related to specific templates for inflationary power spectrum \cite{vicente}, as explained in \cite{albert}, and represents a non-trivial test of our claims. Indeed, this family of curves differs from the others as, generically, $k_p\neq r_m^{-1}$.  Hence, $r_m$ (the scale on which $C'=0$) does {\it not} define a characteristic scale for $K$.  Nevertheless, as we have checked numerically, the associated threshold value still mainly depends on the behaviour of $K$ around its own $r_m$. 

\begin{figure}[h]
	\includegraphics[width=0.8\linewidth]{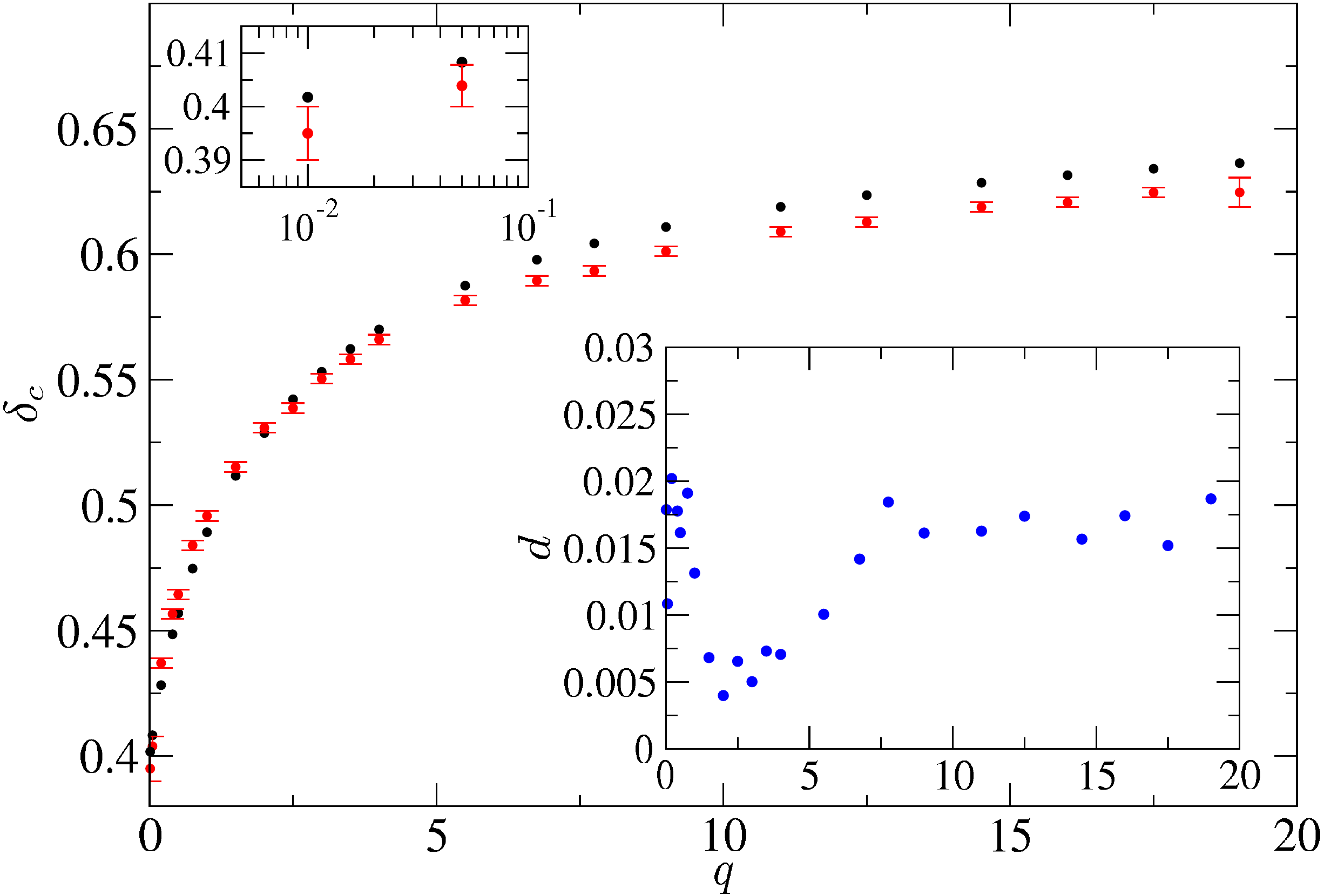} 
	\caption{Threshold values for profiles given by \eqref{eq:pol} for various values of $p$ (red symbols).  For each $p$, $q$ is obtained by fitting the profile shape to \eqref{basis} around $r_m$.  This $q$ is used in \eqref{deltac} to predict the threshold value (black symbols).  As in Fig.~\ref{fig:exponential}, lower inset shows the relative difference between measured and predicted values.  The upper subplot shows that the minimal threshold is below the HYK limit $\delta_c=0.41$. Note that the error bars reflect only the numerical precision of the code.}
	\label{fig:poli}
\end{figure}

\begin{figure}[h]
	\includegraphics[width=0.8\linewidth]{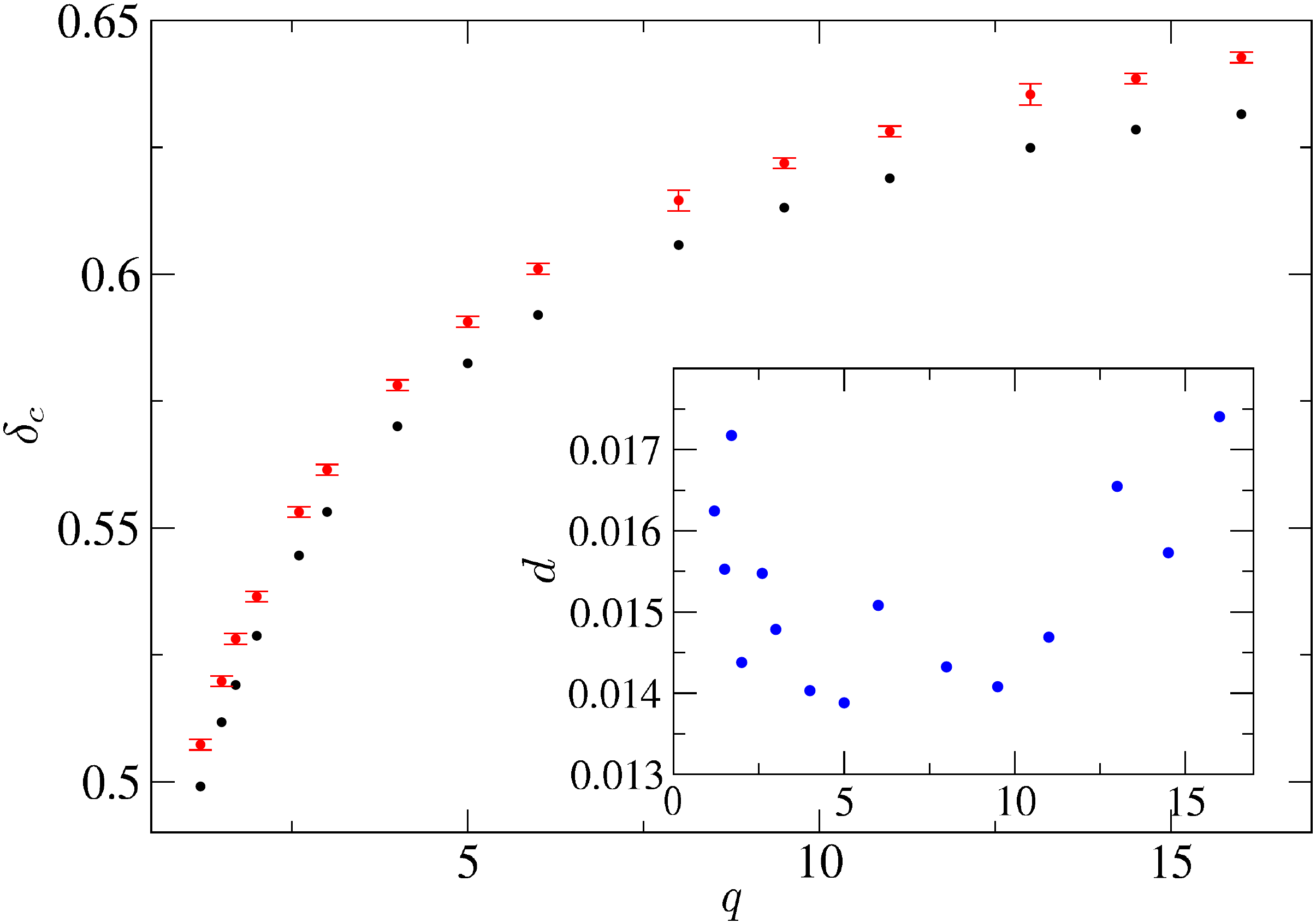} 
	\caption{Same as Fig.~\ref{fig:poli} but for profiles given by \eqref{eq:lamda} with $\alpha=1$ and various $\lambda$.
        }
	\label{fig:lambda}
\end{figure}

\begin{figure}[h]
	\includegraphics[width=0.8\linewidth]{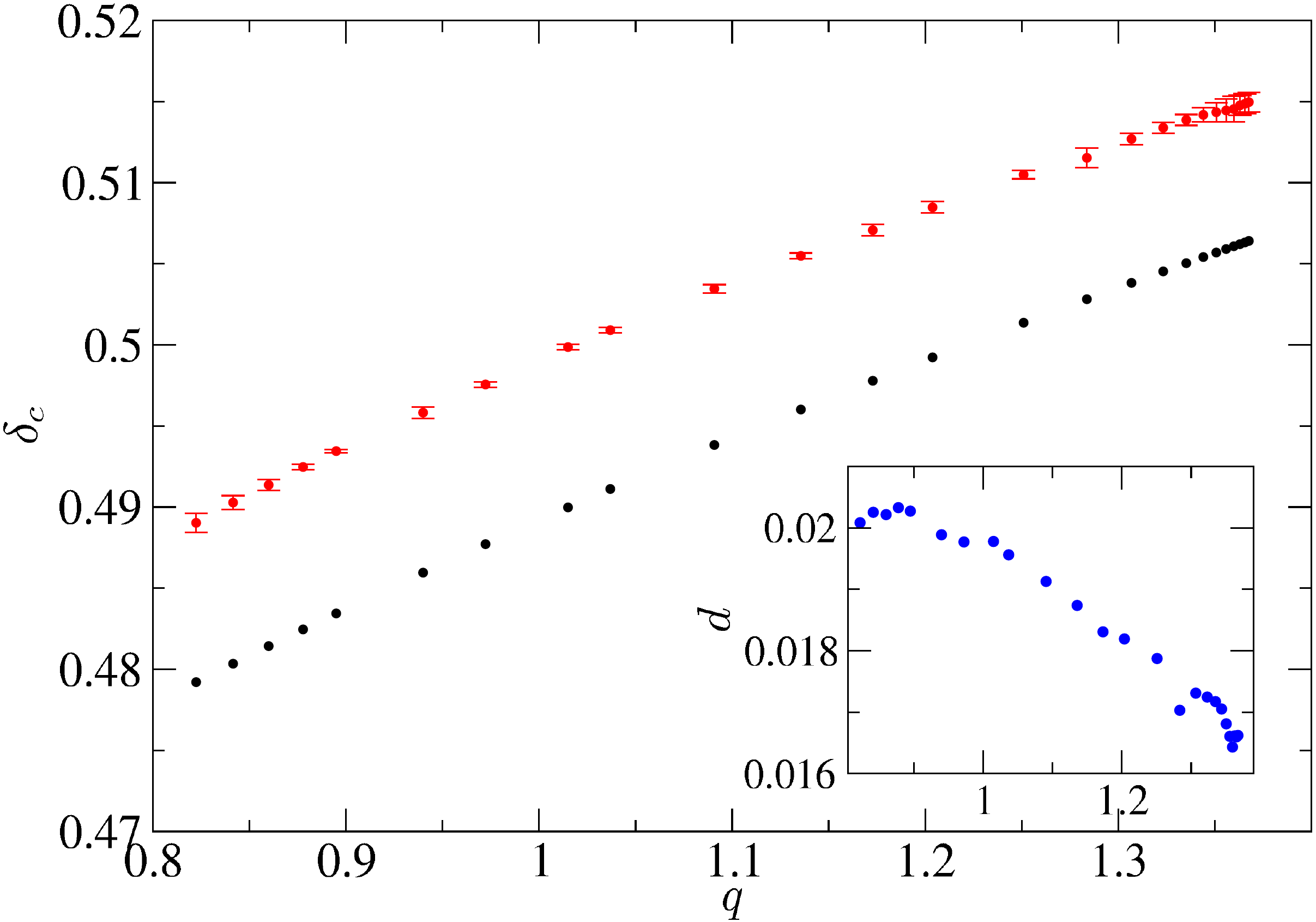} 
	\caption{Same as Fig.~\ref{fig:poli} but for profiles given by \eqref{eq:spectrum}.
        }
	\label{fig:spectrum}
\end{figure}

Figs.~\ref{fig:poli}, \ref{fig:lambda} and \ref{fig:spectrum} are similar in format to Fig.~\ref{fig:exponential}:  they show that \eqref{threshold}, using $q$ calibrated by fitting to $K$, provides a good description of $\delta_{c}^{N}$.  The upper inner plot of Fig.~\ref{fig:poli} shows that $\delta_c^{N}$ reaches values that are slightly smaller than the 0.41 limit quoted in \cite{harada} and later in \cite{musco}, but they do not drop below the 0.4 limit of our Eq.~(\ref{threshold}).

\subsection{Analytical argument}

The threshold for the average compaction function ($\cc=2/5=0.4$) is very close to the so-called Harada-Yoo-Kohri (HYK) limit which was analytically found to be $\sim 0.41$. The value of the second significant digit is related to assumptions about the Jeans length of the perturbation \cite{harada}. As already mentioned, in Fig. \ref{fig:poli}, we show that this limit is actually closer to our theoretical value of $0.40$. Nevertheless, we shall still call the minimal threshold the HYK limit, as the interpretation of it will not change.

The HYK limit is the threshold for which a very sharply peaked over-density profile would collapse into a zero mass black hole, as discussed in \cite{musco}. Let us then approximate an initial over-density to be a Dirac delta function
\be
 \frac{\delta\rho}{\rho}\propto\delta_{\rm D}(\frac{r}{r_m}-1)\ .
\ee
One can always find an initial time where the linear approximation is good enough and find \cite{germani} (see also \cite{sg})
\be
 \com(r)\propto \frac{1}{r}\int_0^r \frac{\delta\rho(x)}{\rho}x^2 dx
\ee
and therefore
\be\label{harada}
 \com(r)=\frac{r_m}{r}\com(r_m)\theta(r-r_m)\ .
\ee
Because, as discussed above, what happens at $r>r_m$ is not crucial for the calculation of the threshold, we can approximate \eqref{harada} as a very thin shell with finite amplitude $\com(r_m)$ positioned at $r=r_m$. In other words we shall cut-off the tail in \eqref{harada}.  The HYK limit indicates that such a shell would collapse and form a zero mass black hole if $\com(r_m)\sim 0.4$.

Now suppose we have a continuum of concentric shells forming a homogeneous ball. This ball would then collapse to a black hole of zero mass if each shell had the same amplitude equal to the HYK threshold.  Our averaging relates the problem of a generic compaction function shape to this homogeneous one.

We then conjecture that the same would happen for any $\omega$ and so, for a generic fluid matter, the threshold would be obtained for
\be
 \cc=\com_{c}^{\rm hom}(\omega)\ ,
\ee 
where $\com_{c}^{\rm hom}$ is the threshold for a homogeneous ball. As a first approximation, one then may be tempted to consider the functional form \cite{harada}
$\cc\sim f(\omega)\sin^2\left[\frac{\pi\sqrt{\omega}}{1+3\omega}\right]$, however, we have numerically checked that, for $\omega\neq 1/3$, the HYK formula for $\com_{c}^{\rm hom}(\omega)$ leads to large errors.

While for radiation we could exactly fix $\cc$ by using the limit of very peaked compaction function, we cannot do the same for other $\omega$. We then leave for future work the extensive proof of our conjecture.

\section{Conclusions}
Primordial black holes can account for the majority of dark matter if they are in the range of $\left[10^{-16},10^{-12}\right]\, M_\odot$ (see e.g. \cite{last}). The seeds for primordial black hole formation might be generated by large statistical fluctuations during inflation. The abundance of these statistical fluctuation, and in turn of the generated PBHs, is extremely sensitive to the threshold required to form a PBH \cite{sg,germani,vicente, jaume}. This threshold, depending upon the full non-linear evolution of the system, is typically given at super-horizon scales where the leading order in gradient expansion is an excellent approximation (see e.g. \cite{albert}). 

To date, analytical estimates of it (see for example \cite{carr,harada}) are insufficiently accurate, so numerical analyses have been employed (for the latest results see \cite{musco,albert}).

In this paper, we have shown that although the threshold to form a PBH is initial curvature profile dependent, as noticed by \cite{musco}, the threshold for the mean (i.e. volume averaged) compaction function within a sphere of radius $r=r_m$, is, to a very good approximation, {\it universal} and equal to the one obtained in the Harada-Yoo-Kohri limit. We used this remarkable result to provide an analytical formula for the threshold that only depends upon the normalised second derivative of the compaction function at its maximum.  Specifically, for a radiation dominated universe, the threshold for a compaction function $\com(r)$ is 
given by Eq.~\ref{deltac}.

\begin{acknowledgments}
AE and CG would like to thank Vicente Atal, Jaume Garriga and Ilia Musco for many illuminating discussions. CG is supported by the Ramon y Cajal program and partially supported by the Unidad de Excelencia Maria de Maeztu Grant No. MDM-2014-0369. AE and CG are partially supported by the national FPA2016-76005-C2-2-P grants. AE is supported by the Spanish MECD fellowship FPU15/03583.
\end{acknowledgments}


\end{document}